\documentclass[twocolumn,letterpaper,aps,prl,longbibliography,superscriptaddress,showpacs,floatfix]{revtex4-1}
\usepackage{graphicx}	% Include figure files

\usepackage{amsmath}	% better math options
\usepackage{xspace}	% Include xspace

\newcommand{\pp}{\mbox{$p$$+$$p$}\xspace}

\newcommand{\dau}{\mbox{$d$$+$Au}\xspace}
\newcommand{\heau}{\mbox{$^{3}$He$+$Au}\xspace}

\newcommand{\pau}{\mbox{$p$$+$Au}\xspace}
\newcommand{\pal}{\mbox{$p$$+$Al}\xspace}
\newcommand{\pa}{\mbox{$p$$+$$A$}\xspace}
\newcommand{\auau}{\mbox{Au$+$Au}\xspace}
\newcommand{\pbpb}{\mbox{Pb$+$Pb}\xspace}

\newcommand{\Nqp}{\mbox{$N_{qp}$}\xspace}

\newcommand{\snn}{$\sqrt{s_{_{NN}}}$}

\newcommand{\dndeta}{dN_{\rm ch}/d\eta}

\begin{document}

\title{Pseudorapidity dependence of particle production and elliptic
flow in asymmetric nuclear collisions of $p$$+$Al, $p$$+$Au, $d$$+$Au,
and $^{3}$He$+$Au at $\sqrt{s_{_{NN}}}=200$ GeV}

\newcommand{\abilene}{Abilene Christian University, Abilene, Texas 79699, USA}
\newcommand{\augie}{Department of Physics, Augustana University, Sioux Falls, South Dakota 57197, USA}
\newcommand{\banaras}{Department of Physics, Banaras Hindu University, Varanasi 221005, India}
\newcommand{\barc}{Bhabha Atomic Research Centre, Bombay 400 085, India}
\newcommand{\baruch}{Baruch College, City University of New York, New York, New York, 10010 USA}
\newcommand{\bnlcoll}{Collider-Accelerator Department, Brookhaven National Laboratory, Upton, New York 11973-5000, USA}
\newcommand{\bnlphys}{Physics Department, Brookhaven National Laboratory, Upton, New York 11973-5000, USA}
\newcommand{\caucr}{University of California-Riverside, Riverside, California 92521, USA}
\newcommand{\charlesczech}{Charles University, Ovocn\'{y} trh 5, Praha 1, 116 36, Prague, Czech Republic}
\newcommand{\chonbuk}{Chonbuk National University, Jeonju, 561-756, Korea}
\newcommand{\ciae}{Science and Technology on Nuclear Data Laboratory, China Institute of Atomic Energy, Beijing 102413, People's Republic of China}
\newcommand{\cns}{Center for Nuclear Study, Graduate School of Science, University of Tokyo, 7-3-1 Hongo, Bunkyo, Tokyo 113-0033, Japan}
\newcommand{\colorado}{University of Colorado, Boulder, Colorado 80309, USA}
\newcommand{\columbia}{Columbia University, New York, New York 10027 and Nevis Laboratories, Irvington, New York 10533, USA}
\newcommand{\czechtech}{Czech Technical University, Zikova 4, 166 36 Prague 6, Czech Republic}
\newcommand{\debrecen}{Debrecen University, H-4010 Debrecen, Egyetem t{\'e}r 1, Hungary}
\newcommand{\elte}{ELTE, E{\"o}tv{\"o}s Lor{\'a}nd University, H-1117 Budapest, P{\'a}zm{\'a}ny P.~s.~1/A, Hungary}
\newcommand{\eszterhazy}{Eszterh\'azy K\'aroly University, K\'aroly R\'obert Campus, H-3200 Gy\"ongy\"os, M\'atrai \'ut 36, Hungary}
\newcommand{\ewha}{Ewha Womans University, Seoul 120-750, Korea}
\newcommand{\fsu}{Florida State University, Tallahassee, Florida 32306, USA}
\newcommand{\gsu}{Georgia State University, Atlanta, Georgia 30303, USA}
\newcommand{\hiroshima}{Hiroshima University, Kagamiyama, Higashi-Hiroshima 739-8526, Japan}
\newcommand{\howard}{Department of Physics and Astronomy, Howard University, Washington, DC 20059, USA}
\newcommand{\ihepprot}{IHEP Protvino, State Research Center of Russian Federation, Institute for High Energy Physics, Protvino, 142281, Russia}
\newcommand{\illuiuc}{University of Illinois at Urbana-Champaign, Urbana, Illinois 61801, USA}
\newcommand{\inrras}{Institute for Nuclear Research of the Russian Academy of Sciences, prospekt 60-letiya Oktyabrya 7a, Moscow 117312, Russia}
\newcommand{\instpasczech}{Institute of Physics, Academy of Sciences of the Czech Republic, Na Slovance 2, 182 21 Prague 8, Czech Republic}
\newcommand{\isu}{Iowa State University, Ames, Iowa 50011, USA}
\newcommand{\jaea}{Advanced Science Research Center, Japan Atomic Energy Agency, 2-4 Shirakata Shirane, Tokai-mura, Naka-gun, Ibaraki-ken 319-1195, Japan}
\newcommand{\jyvaskyla}{Helsinki Institute of Physics and University of Jyv{\"a}skyl{\"a}, P.O.Box 35, FI-40014 Jyv{\"a}skyl{\"a}, Finland}
\newcommand{\kek}{KEK, High Energy Accelerator Research Organization, Tsukuba, Ibaraki 305-0801, Japan}
\newcommand{\korea}{Korea University, Seoul, 02841, Korea}
\newcommand{\kurchatov}{National Research Center ``Kurchatov Institute", Moscow, 123098 Russia}
\newcommand{\kyoto}{Kyoto University, Kyoto 606-8502, Japan}
\newcommand{\lawllnl}{Lawrence Livermore National Laboratory, Livermore, California 94550, USA}
\newcommand{\losalamos}{Los Alamos National Laboratory, Los Alamos, New Mexico 87545, USA}
\newcommand{\lund}{Department of Physics, Lund University, Box 118, SE-221 00 Lund, Sweden}
\newcommand{\lyon}{IPNL, CNRS/IN2P3, Univ Lyon, Université Lyon 1, F-69622, Villeurbanne, France}
\newcommand{\maryland}{University of Maryland, College Park, Maryland 20742, USA}
\newcommand{\mass}{Department of Physics, University of Massachusetts, Amherst, Massachusetts 01003-9337, USA}
\newcommand{\michigan}{Department of Physics, University of Michigan, Ann Arbor, Michigan 48109-1040, USA}
\newcommand{\muhlenberg}{Muhlenberg College, Allentown, Pennsylvania 18104-5586, USA}
\newcommand{\nara}{Nara Women's University, Kita-uoya Nishi-machi Nara 630-8506, Japan}
\newcommand{\natmephi}{National Research Nuclear University, MEPhI, Moscow Engineering Physics Institute, Moscow, 115409, Russia}
\newcommand{\newmex}{University of New Mexico, Albuquerque, New Mexico 87131, USA}
\newcommand{\nmsu}{New Mexico State University, Las Cruces, New Mexico 88003, USA}
\newcommand{\ohio}{Department of Physics and Astronomy, Ohio University, Athens, Ohio 45701, USA}
\newcommand{\ornl}{Oak Ridge National Laboratory, Oak Ridge, Tennessee 37831, USA}
\newcommand{\orsay}{IPN-Orsay, Univ.~Paris-Sud, CNRS/IN2P3, Universit\'e Paris-Saclay, BP1, F-91406, Orsay, France}
\newcommand{\peking}{Peking University, Beijing 100871, People's Republic of China}
\newcommand{\pnpi}{PNPI, Petersburg Nuclear Physics Institute, Gatchina, Leningrad region, 188300, Russia}
\newcommand{\riken}{RIKEN Nishina Center for Accelerator-Based Science, Wako, Saitama 351-0198, Japan}
\newcommand{\rikjrbrc}{RIKEN BNL Research Center, Brookhaven National Laboratory, Upton, New York 11973-5000, USA}
\newcommand{\rikkyo}{Physics Department, Rikkyo University, 3-34-1 Nishi-Ikebukuro, Toshima, Tokyo 171-8501, Japan}
\newcommand{\saispbstu}{Saint Petersburg State Polytechnic University, St.~Petersburg, 195251 Russia}
\newcommand{\seoulnat}{Department of Physics and Astronomy, Seoul National University, Seoul 151-742, Korea}
\newcommand{\stonybrkc}{Chemistry Department, Stony Brook University, SUNY, Stony Brook, New York 11794-3400, USA}
\newcommand{\stonycrkp}{Department of Physics and Astronomy, Stony Brook University, SUNY, Stony Brook, New York 11794-3800, USA}
\newcommand{\tenn}{University of Tennessee, Knoxville, Tennessee 37996, USA}
\newcommand{\titech}{Department of Physics, Tokyo Institute of Technology, Oh-okayama, Meguro, Tokyo 152-8551, Japan}
\newcommand{\tsukuba}{Tomonaga Center for the History of the Universe, University of Tsukuba, Tsukuba, Ibaraki 305, Japan}
\newcommand{\vandy}{Vanderbilt University, Nashville, Tennessee 37235, USA}
\newcommand{\weizmann}{Weizmann Institute, Rehovot 76100, Israel}
\newcommand{\wigner}{Institute for Particle and Nuclear Physics, Wigner Research Centre for Physics, Hungarian Academy of Sciences (Wigner RCP, RMKI) H-1525 Budapest 114, POBox 49, Budapest, Hungary}
\newcommand{\yonsei}{Yonsei University, IPAP, Seoul 120-749, Korea}
\newcommand{\zagreb}{Department of Physics, Faculty of Science, University of Zagreb, Bijeni\v{c}ka c.~32 HR-10002 Zagreb, Croatia}
\affiliation{\abilene}
\affiliation{\augie}
\affiliation{\banaras}
\affiliation{\barc}
\affiliation{\baruch}
\affiliation{\bnlcoll}
\affiliation{\bnlphys}
\affiliation{\caucr}
\affiliation{\charlesczech}
\affiliation{\chonbuk}
\affiliation{\ciae}
\affiliation{\cns}
\affiliation{\colorado}
\affiliation{\columbia}
\affiliation{\czechtech}
\affiliation{\debrecen}
\affiliation{\elte}
\affiliation{\eszterhazy}
\affiliation{\ewha}
\affiliation{\fsu}
\affiliation{\gsu}
\affiliation{\hiroshima}
\affiliation{\howard}
\affiliation{\ihepprot}
\affiliation{\illuiuc}
\affiliation{\inrras}
\affiliation{\instpasczech}
\affiliation{\isu}
\affiliation{\jaea}
\affiliation{\jyvaskyla}
\affiliation{\kek}
\affiliation{\korea}
\affiliation{\kurchatov}
\affiliation{\kyoto}
\affiliation{\lawllnl}
\affiliation{\losalamos}
\affiliation{\lund}
\affiliation{\lyon}
\affiliation{\maryland}
\affiliation{\mass}
\affiliation{\michigan}
\affiliation{\muhlenberg}
\affiliation{\nara}
\affiliation{\natmephi}
\affiliation{\newmex}
\affiliation{\nmsu}
\affiliation{\ohio}
\affiliation{\ornl}
\affiliation{\orsay}
\affiliation{\peking}
\affiliation{\pnpi}
\affiliation{\riken}
\affiliation{\rikjrbrc}
\affiliation{\rikkyo}
\affiliation{\saispbstu}
\affiliation{\seoulnat}
\affiliation{\stonybrkc}
\affiliation{\stonycrkp}
\affiliation{\tenn}
\affiliation{\titech}
\affiliation{\tsukuba}
\affiliation{\vandy}
\affiliation{\weizmann}
\affiliation{\wigner}
\affiliation{\yonsei}
\affiliation{\zagreb}
\author{A.~Adare} \affiliation{\colorado}
\author{C.~Aidala} \affiliation{\michigan}
\author{N.N.~Ajitanand} \altaffiliation{Deceased} \affiliation{\stonybrkc}
\author{Y.~Akiba} \email[PHENIX Spokesperson: ]{akiba@rcf.rhic.bnl.gov} \affiliation{\riken} \affiliation{\rikjrbrc}
\author{M.~Alfred} \affiliation{\howard}
\author{V.~Andrieux} \affiliation{\michigan}
\author{K.~Aoki} \affiliation{\kek} \affiliation{\riken}
\author{N.~Apadula} \affiliation{\isu} \affiliation{\stonycrkp}
\author{H.~Asano} \affiliation{\kyoto} \affiliation{\riken}
\author{C.~Ayuso} \affiliation{\michigan}
% ...
\collaboration{PHENIX Collaboration} \noaffiliation

\date{\today}

%------------------------------------------------------------------------------|

\begin{abstract}

Asymmetric nuclear collisions of $p$$+$Al, $p$$+$Au, $d$$+$Au, and
$^{3}$He$+$Au at $\sqrt{s_{_{NN}}}=200$ GeV provide an excellent
laboratory for understanding particle production, as well as exploring
interactions among these particles after their initial creation in the
collision.  We present measurements of charged hadron production
$dN_{\rm ch}/d\eta$ in all such collision systems over a broad pseudorapidity
range and as a function of collision multiplicity.  A simple wounded
quark model is remarkably successful at describing the full data set.
We also measure the elliptic flow $v_{2}$ over a similarly broad
pseudorapidity range.  These measurements provide key constraints on
models of particle emission and their translation into flow.

\end{abstract}

\maketitle

%\textbf{*** page break for PRL word count ***}  \clearpage

%%%%%%%%%%%%%%%%%%%%%%% Introduction %%%%%%%%%%%%%%%%%%%%%%%

Asymmetric nuclear collisions with a light projectile nucleus striking a 
heavier target nucleus have proven to be an excellent testing ground for 
particle production models and the longitudinal dynamics following the 
initial collision -- for an early review see Ref.~\cite{Busza:1975te}.  
Many calculations have successfully described the longitudinal (or 
rapidity) distribution of produced particles in proton-nucleus (\pa) 
collisions via the fragmentation of color strings and with counting 
rules based on the number of ``wounded'' or struck nucleons or quarks in 
the projectile and target. Recently, a proposal for testing the 
wounded-quark model~\cite{VoloshinQuarks} was put forth that 
specifically called for the measurement of $\dndeta$ over a broad range 
of pseudorapidity in \pau, \dau, and \heau 
collisions~\cite{Barej:2017kcw}. Fully three-dimensional hydrodynamical 
models also require input on the longitudinal distribution of initial 
deposited energy and gradients thereof~\cite{BOZEK2014308}. Once the 
initial partons or fluid elements are populated, the models evolve the 
system dynamically. Measurements of elliptic flow as a function of 
pseudorapidity provide constraints on the longitudinal dynamics of the 
evolution.

As the incoming hadrons or nuclei break up, the rapidity distribution of 
liberated partons may be determined by the longitudinal parton 
distribution functions~\cite{PhysRev.188.2159,Gelis:2006tb} or via a 
universal color field breakup for each struck nucleon or 
quark~\cite{Bozek:2016kpf}.  For that reason, calculations based on 
Monte Carlo Glauber models have been developed to calculate the number 
of struck nucleons and struck quarks (see for example 
Refs.~\cite{PhysRevC.89.044905,Mitchell:2016jio,Loizides:2016djv}). The 
PHOBOS collaboration has previously published charged hadron $\dndeta$ 
measurements over $|\eta| < 5.4$ in \dau collisions at 
\snn~=~200~GeV~\cite{Back:2004mr}.  PHENIX has also published $\dndeta$ 
measurements in high-multiplicity \dau collisions at \snn~=~200, 62, 39, 
and 19.6~GeV~\cite{Aidala:2017pup}. The wounded-quark model has been 
constrained by the \dau data and found to be in reasonable agreement 
with the centrality dependence, while the wounded-nucleon model cannot 
describe the data~\cite{Barej:2017kcw}.  A crucial test of the 
wounded-quark model is to see if it is universal across different 
colliding systems. Additional measurements in light and heavy systems at 
the Relativistic Heavy Ion Collider (RHIC) and the Large Hadron Collider 
(LHC) can also be tested in this context---see for example different 
geometry tests in Refs.~\cite{Acharya:2018hhy,STARUUAuAu,Adare:2015bua}.

In \auau and \pbpb collisions at RHIC and the LHC, the created medium is
well described by low viscosity
hydrodynamics~\cite{Heinz:2013th,Romatschke:2017ejr}.  A host of
recent experimental observations indicate that hydrodynamics may also be
applicable to the asymmetric collisions of small nuclear systems, e.g.
\pa, \dau, \heau, and perhaps even \pp (for a recent review see
Ref.~\cite{Nagle:2018nvi}).  In heavy ion collisions, the hydrodynamical
flow of the medium is characterized via a Fourier decomposition of the
final hadron momentum anisotropy in the direction transverse to the
incoming beam directions~\cite{Voloshin:1994mz} as
%\begin{linenomath}
\begin{equation}
\frac{dN}{d\phi} \propto 1 + \sum_n 2 v_n \cos \left[ n \left(\phi - \psi_{n} \right) \right],
\end{equation}
%\end{linenomath}
where $n$ is the harmonic number, $\phi$ is the particle azimuthal 
angle, $\psi_{n}$ is the $n^{th}$ order symmetry axis, and $v_n$ is the 
Fourier coefficient, with $v_{2}$ referred to as elliptic flow. The 
pseudorapidity dependence of $v_{2}$ has been measured in \auau and 
\pbpb collisions at RHIC and the LHC, and the elliptic flow is smaller 
in regions with smaller final hadron $\dndeta$ -- see for example 
Refs.~\cite{Back:2004mh,ALICEPbPbv2eta}. The data have been interpreted 
in terms of hydrodynamics and imply a shear viscosity to entropy 
density, $\eta/s$, that is temperature dependent~\cite{Denicol:2015nhu}.  
Similar measurements in small nuclear collisions of different sizes are 
a key test for how local rapidity density relates to hydrodynamical 
evolution into flow.

In this Letter, we present a comprehensive set of measurements of 
$\dndeta$ and elliptic flow $v_{2}$ over a broad pseudorapidity range in 
\pal, \pau, \dau, and \heau collisions at \snn~=~200~GeV. The data sets 
analyzed were recorded in 2014 for \heau, 2015 for \pal and \pau, and 
2016 for \dau. All data sets were recorded with a minimum-bias trigger 
that required at least one hit in each of the PHENIX beam-beam counters 
(BBC). The BBC is composed of two detectors each containing 64 quartz 
radiators read out with photomultiplier tubes~\cite{Allen:2003zt}. The 
BBC covers positive and negative pseudorapidity $3.1 < |\eta| < 3.9$.  
Following the procedure from Ref.~\cite{Adare:2013nff}, the minimum-bias 
trigger is determined to fire on 88 $\pm$ 4\%, 88 $\pm$ 4\%, 84 $\pm$ 
3\%, and 72 $\pm$ 4\% of the total inelastic cross section of 2.30, 
2.26, 1.76, 0.54 barns for \heau, \dau, \pau, and \pal respectively. The 
$\dndeta$ analysis has negligible statistical uncertainties and thus a 
subset of runs with the most stable detector configuration are utilized 
and the run-to-run variation is used in the determination of systematic 
uncertainties.  For the elliptic flow $v_{2}$ analysis in 
high-multiplicity events, also referred to as central events, an 
additional trigger was used that required the number of fired BBC tubes 
to be above a set number, roughly corresponding to the 0\%--5\% highest 
multiplicity events.

The characterization of the different collision systems and centralities
follows the procedure detailed in Ref.~\cite{Adare:2013nff}.  The
multiplicity class is selected by the total charge in the BBC covering
negative pseudorapidity, i.e. in the Al- or Au-going direction.  The
total charge is found to scale with the total number of struck nucleons
from the Al or Au nucleus folded with a negative binomial distribution
representing the fluctuations in the number of particles produced and
measured by the BBC.  The 5\% most central events have an average number
of participating nucleons of 5.1 $\pm$ 0.3, 10.7 $\pm$ 0.6, 17.8 $\pm$
1.2, and 25.0 $\pm$ 1.6 for \pal, \pau, \dau, and \heau respectively.

Charged hadrons are reconstructed at midrapidity $|\eta| < 0.35$ with a
combination of drift chambers and pad chambers~\cite{Adcox:2003zp}.
Midrapidity tracks have their momentum reconstructed via their bend in a
magnetic field and are efficiently measured for $p_{T} >$ 0.2~GeV/$c$.
At backward $-3.0 < \eta < -1.0$ and forward $1.0 < \eta < 3.0$
rapidity, the forward-silicon-vertex detector (FVTX) measures the
traversal of charged tracks in four detector layers as detailed in
Ref.~\cite{Aidala:2013vna}. FVTX tracks are efficiently measured for
$p_{T} >$ 0.3~GeV/$c$, but with no momentum information, because the
silicon strips are oriented lengthwise along the magnetic field bend
direction.

For the $\dndeta$ results, the absolute acceptance and efficiency for 
track reconstruction can be determined with the PHENIX {\sc geant}-3 
Monte Carlo simulation.  However, in the last years of data taking, the 
PHENIX experiment had increasingly significant dead regions and 
run-to-run variations that became challenging to fully account for. 
Thus, we determine the acceptance and efficiency for a given running 
period in a control data set by taking the ratio $R(\eta)$ of published 
PHOBOS $\dndeta$ to the PHENIX raw $\dndeta$ as a function of 
pseudorapidity.  The control PHOBOS data sets are \auau in 
2014~\cite{Alver:2010ck}, \pp in 2015~\cite{Alver:2010ck}, and \dau in 
2016~\cite{Back:2004mr} all at \snn~=~200~GeV.  This ``bootstrapping'' 
procedure is described in detail in Ref.~\cite{Aidala:2017pup}. Sources 
of systematic uncertainty come from varying the track selection cuts, 
run-to-run variations, and considering high and low luminosity running 
periods with different double interaction contributions.  We also find 
good agreement within uncertainties comparing results in the FVTX with 
an absolute acceptance and efficiency calculation and the 
``bootstrapped'' results.

The determination of hadron yields in centrality bins has a known bias 
effect (see Ref.~\cite{Adare:2013nff}). In \pp collisions, inelastic 
events fire the BBC trigger 55 $\pm$ 5\% of the time, while in events 
with a $\pi^{0}$ or charged hadron at midrapidity that percentage is 
larger, 79 $\pm$ 2\%. This increased trigger efficiency is correlated 
with a 1.55 times larger BBC multiplicity.  This effect results from the 
diffractive portion of the \pp inelastic cross section disfavoring 
midrapidity particle production. This bias has been confirmed for 
midrapidity hadron production down to $p_{T} \approx 
0.5$~GeV/$c$~\cite{Adler:2005in} and for $J/\psi$ measured in the PHENIX 
muons arms~\cite{Adler:2005ph}, and thus we expect that this bias 
affects all charged hadrons over the pseudorapidity range studied here.  
We remove this bias via correction factors that are calculated following 
the procedure detailed in Ref.~\cite{Adare:2013nff}. The bias 
corrections are largest in the smallest system and range from 0.75 $\pm$ 
0.01 for central 0\%--5\% \pal to 0.91 $\pm$ 0.01 for central 0\%--5\% 
\heau. We apply these bias correction factors to all our $\dndeta$ 
results.

%%%%%%%%%%%%%%%%%%%%%%%%%%%%%%%%%%%%%%%%%%%%%%%%%%%%%%%%%%%%%% Fig_1
\begin{figure}[hbt]
\includegraphics[width=1.0\linewidth]{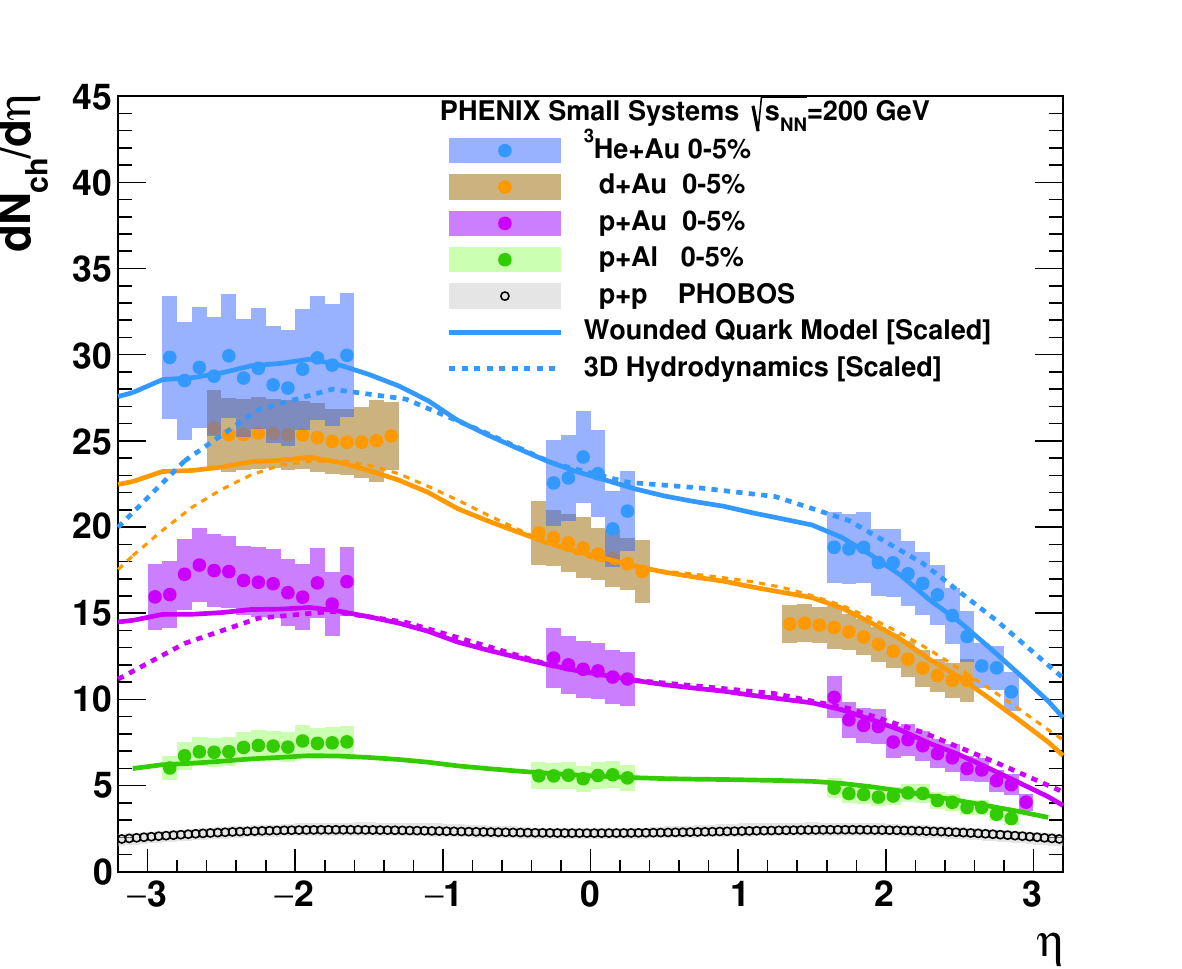}
\caption{
Charged hadron $\dndeta$ as a function of pseudorapidity in 
high-multiplicity 0\%--5\% central \heau, \dau, \pau, and \pal 
collisions at \snn~=~200~GeV.  Also shown are results in inelastic \pp 
collisions at \snn~=~200~GeV as measured by the PHOBOS 
Collaboration~\cite{Alver:2010ck}.  Predictions from the 
wounded-quark~\cite{Barej:2017kcw} and 
hydrodynamical~\cite{BOZEK2014308} models are shown.  The calculations 
have an overall normalization factor ($S$) to best match the data.  
These factors are $S$=0.88, 0.93, 0.85, 0.77 for the wound quark model 
for \pal, \pau, \dau, \heau respectively, and $S$=0.81, 0.96, 0.75 for 
the hydrodynamical model for \pau, \dau, \heau respectively.
}
\label{fig:dnchdeta_central}
\end{figure}

%%%%%%%%%%%%%%%%%%%%%%%%%%%%%%%%%%%%%%%%%%%%%%%%%%%%%%%%%%%%%% Fig_2
\begin{figure*}[thb]
\includegraphics[width=0.99\linewidth]{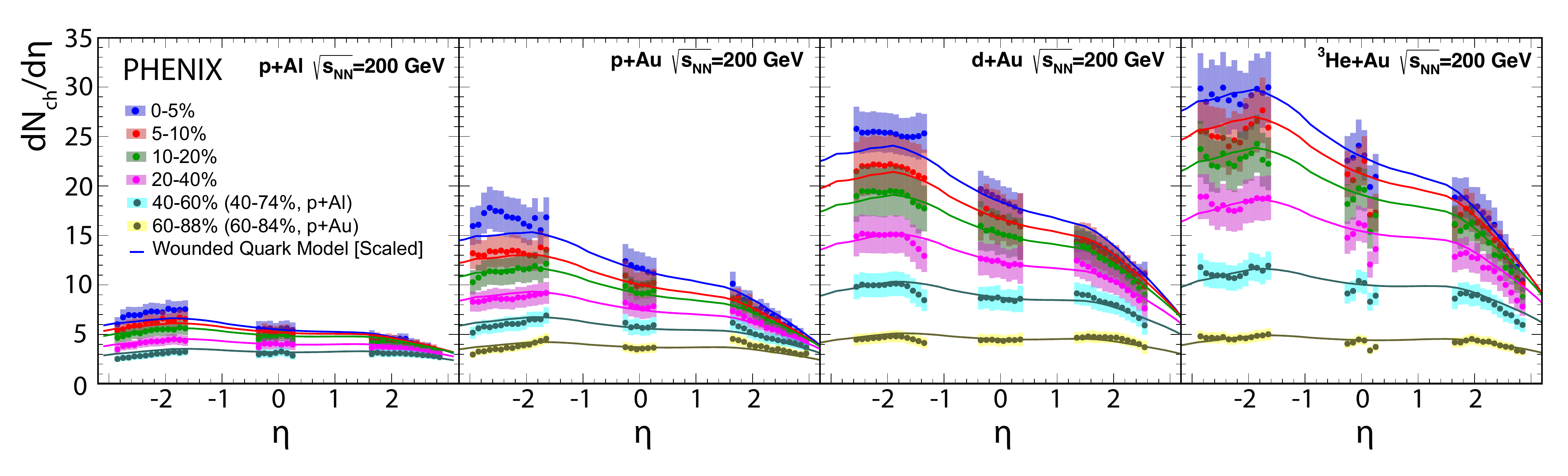}
\caption{
Charged hadron $\dndeta$ as a function of pseudorapidity in various
multiplicity classes of \pal, \pau, \dau, \heau collisions at
\snn~=~200~GeV.  Predictions from the wounded-quark
model~\cite{Barej:2017kcw} are shown.
}
\label{fig:dnchdeta_all}
\end{figure*}

%%%%%%%%%%%%%%%%%%%%%%%%%%%%%%%%%%%%%%%%%%%%%%%%%%%%%%%%%%%%%%%  Fig_3
\begin{figure}[hb]
\includegraphics[width=1.0\linewidth]{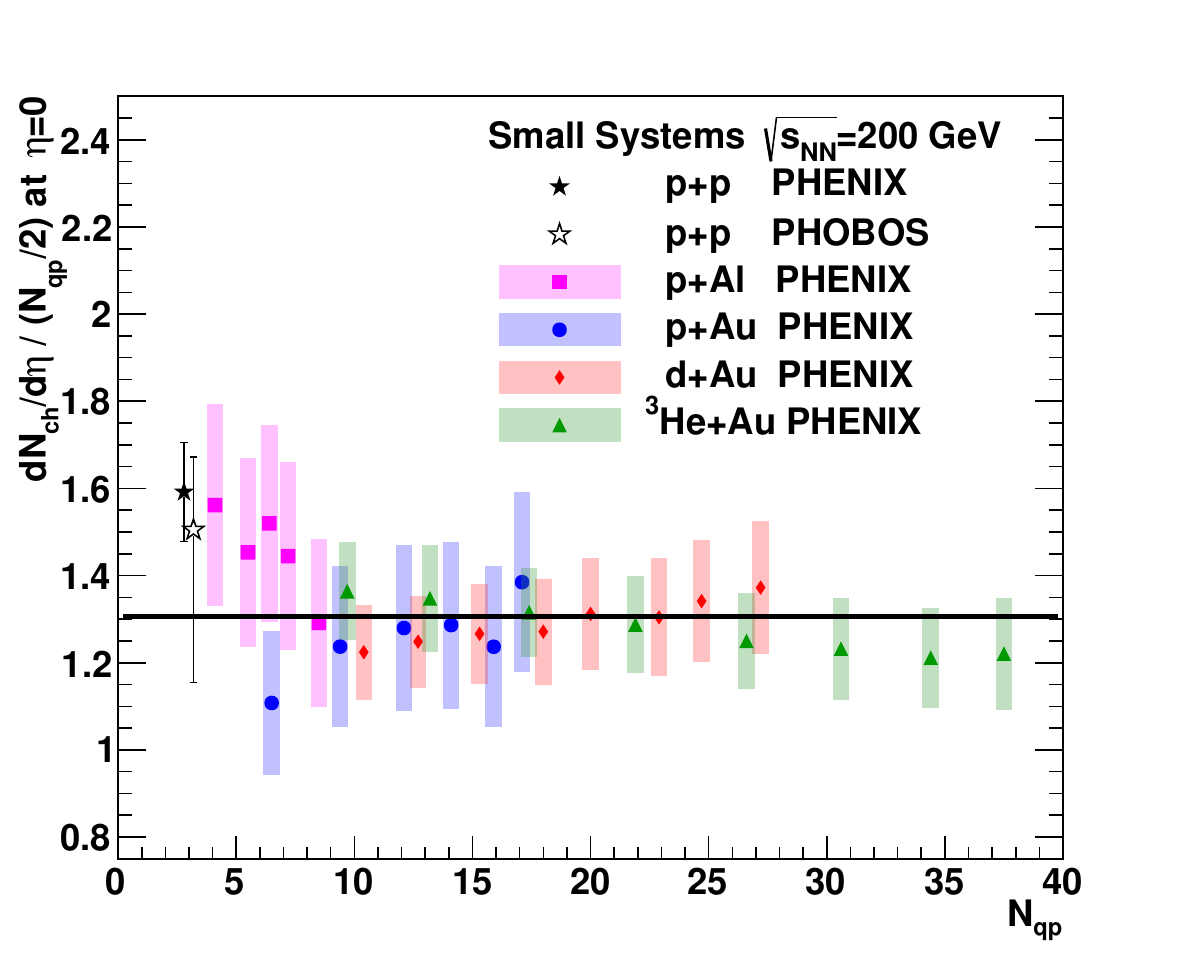}
\caption{
Midrapidity charged hadron $\dndeta$ per participating quark pair
($\Nqp$/2) as a function of the number of participating quarks ($\Nqp$).
Results are shown for \pal, \pau, \dau, and \heau collisions in various
multiplicity classes.  Also shown are previously published results in
\pp collisions from PHENIX~\cite{Adare:2015bua} and
PHOBOS~\cite{Alver:2010ck}.  The line is the best fit to all the data to
a constant level.
  }
\label{fig:npartscaling}
\end{figure}

Figure~\ref{fig:dnchdeta_central} shows the $\dndeta$ results
for \pal, \pau, \dau, and \heau at \snn~=~200~GeV for the 5\% highest
multiplicity events.  Statistical uncertainties are
negligible and systematic uncertainties are shown as boxes around the
points.  The systematic uncertainties are point-to-point correlated and
can in principle move the backward, mid, and forward rapidity points
separately because they are measured in different detectors. Also shown
are the yields in inelastic \pp collisions at \snn~=~200~GeV as measured
by the PHOBOS Collaboration~\cite{Alver:2010ck}.  The full set of
multiplicity-selected results for the four asymmetric nuclear collision
systems are shown in Fig.~\ref{fig:dnchdeta_all}.

The results are compared to predictions from the wounded-quark model. 
Within the wounded-quark model, each wounded-quark is posited to yield 
hadrons following a common emission function 
$F(\eta)$~\cite{Barej:2017kcw}.  $F(\eta)$ is constrained by \dau 
collision data, and the model then predicts $\dndeta$ for all collision 
centralities and systems. The calculations are normalized, with factors 
listed in the Fig.~\ref{fig:dnchdeta_central} caption, to best match the 
data integrated over pseudorapidity, because the exact normalization can 
be influenced by modest differences in the centrality selection and thus 
the mean number of wounded quarks.  Within the systematic uncertainties 
on the experimental measurements, the model provides a good description 
of the complete data set across collision systems and centrality 
classes. The results are also compared in 
Fig.~\ref{fig:dnchdeta_central} with a hydrodynamical 
calculation~\cite{BOZEK2014308} for 0\%--5\% central collisions.  The 
calculation includes Monte Carlo Glauber initial conditions with 
longitudinal entropy distributions~\cite{Bozek:2012fw}, 3+1D viscous 
hydrodynamics~\cite{Bozek:2011ua} with $\eta/s = 1/4\pi$ and temperature 
dependent bulk viscosity, followed by statistical hadronization.  Again, 
the calculations are normalized to the data with factors listed in the 
caption.  The agreement in this case is also good within systematic 
uncertainties, except for a more significant drop in particle yield in 
the calculation at the most backward rapidity region $-3.0 < \eta 
\lesssim -2.0$.

Midrapidity $\dndeta$ per participating quark pair, $\Nqp/2$, scales as
a function of the number of participating quarks from \dau and \heau
collisions~\cite{Adare:2015bua}.  The previously reported
results~\cite{Adare:2015bua} were not corrected for the modest bias
previously discussed.  Figure~\ref{fig:npartscaling} shows the
results testing this scaling for all small collision systems, each
with the bias correction factors applied.  Within the systematic
uncertainties, all systems at all centralities follow a common scaling
for midrapidity particle production.

In \dau collisions, the elliptic flow $v_{2}$ was observed to have a
similar pseudorapidity dependence as the particle yield
$\dndeta$~\cite{Aidala:2017pup}. For the other systems we have followed
the same procedure for measuring elliptic flow $v_{2}$ using the event
plane method, where the event plane is defined by the Al- or Au-going
BBC covering $-3.9 < \eta < -3.1$.  The results are corrected using {\sc
ampt}~\cite{Lin:2004en} and a {\sc geant}-3 simulation of the detector
to correspond to $v_{2}$ integrated over hadrons at all $p_{T}$ within
each pseudorapidity bin.  Systematic uncertainties are determined by
varying the track selection cuts, collision z-vertex cuts, and {\sc
ampt} input parameters.

%%%%%%%%%%%%%%%%%%%%%%%%%%%%%%%%%%%%%%%%%%%%%%%%%%%%%%%%%%%%%%%  Fig_4
\begin{figure*}[thb]
\includegraphics[width=0.99\linewidth]{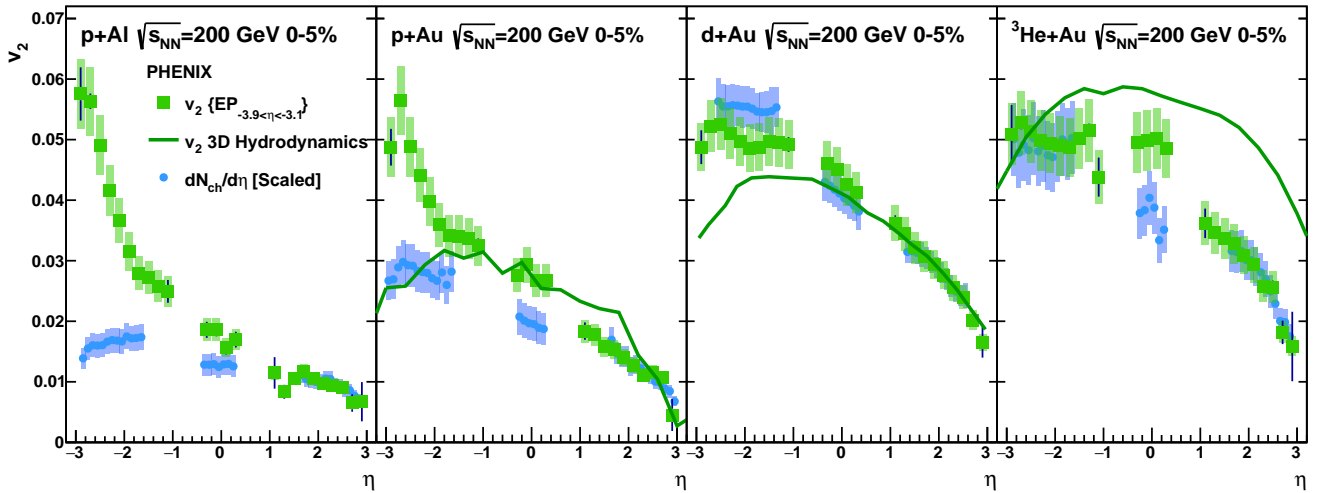}
\caption{
Elliptic flow $v_{2}$ as a function of pseudorapidity in
high-multiplicity 0\%--5\% central \pal, \pau, \dau, and \heau collisions
at \snn~=~200~GeV.  Also shown are predictions from the hydrodynamical
model~\cite{BOZEK2014308}.  Lastly, the measured $\dndeta$ results are
shown scaled to match the $v_{2}$ at forward rapidity for shape
comparison with the elliptic flow coefficients.
  }
\label{fig:v2scaling}
\end{figure*}

Figure~\ref{fig:v2scaling} shows the elliptic flow $v_{2}$ as a function
of pseudorapidity in 0\%--5\% central \pal, \pau, \dau, and \heau
collisions at \snn~=~200~GeV.  The experimental data have an increasing
flow coefficient at forward rapidity when going from the smallest system
and smallest particle production \pal to the largest \heau. These trends
are consistent with arising from the combined influence of initial
geometry and particle multiplicity~\cite{Aidala:2018mcw}. The $v_{2}$
also increases towards backward rapidity for each collision system.  For
the lowest multiplicity systems \pal and \pau, there is a sharp
enhancement in the $v_{2}$ for $\eta \lesssim -2.0$ that is more
pronounced in \pal.  This feature may be due to the nonflow contribution
of short range correlations, because this is the pseudorapidity range that
is within one unit of the BBC used for determining the event plane.

The data are compared with the same hydrodynamical
model~\cite{BOZEK2014308} that gave a reasonable description of the
$\dndeta$.  There is good qualitative agreement with the system and
pseudorapidity dependence of $v_{2}$, and good quantitative agreement of
its pseudorapidity dependence in \pau and \dau.  The only feature not
qualitatively described is the enhancement at backward rapidity.  This
enhancement is the strongest in \pal, weaker but still pronounced in
\pau, and rather weak in \dau.  The strength of this enhancement trends
inversely with the $\dndeta$, lending additional evidence that this is
due to nonflow influences not incorporated in the hydrodynamical model.
In \heau collisions, the hydrodynamical model overpredicts the forward
rapidity ($\eta>$~1) $v_2$ by more than 50\% and qualitatively has the
feature of a weaker forward/backward asymmetry than what is present in
the data.  Note that the model overpredicts the \heau $\dndeta$
by approximately 25\% (but is scaled to fit the data in
Fig.~\ref{fig:dnchdeta_central}), which may help explain
the overpredicted $v_2$.

In Fig.~\ref{fig:v2scaling}, we also scale $\dndeta$ to match the
$v_{2}$ at forward rapidity to compare the shape of the distributions.
Although a larger local particle density $\dndeta$ is correlated with
more elliptic flow, the scaling observed in \dau appears only
approximate when viewed in the context of all collision systems.  It is
notable that although not shown in Fig.~\ref{fig:v2scaling},
hydrodynamical model calculations~\cite{BOZEK2014308} also do not
exhibit an exact scaling relation $v_{2} \propto \dndeta$.

%%%%%%%%%%%%%%%%%%%
%%% now the summary
%%%%%%%%%%%%%%%%%%%

We have presented a comprehensive set of measurements of particle 
production $\dndeta$ and elliptic flow $v_{2}$ over a broad 
pseudorapidity range for a suite of asymmetric nuclear collisions \pal, 
\pau, \dau, and \heau at \snn~=~200~GeV.  The particle production is 
remarkably well-described in the context of the wounded-quark 
model~\cite{Barej:2017kcw}. A three-dimensional hydrodynamical model 
qualitatively describes the particle production and elliptic flow in 
high-multiplicity events in all collision systems. However, it over 
predicts the overall $\dndeta$ and forward rapidity $v_{2}$ in \heau 
collisions. These data provide an important constraint on models of the 
longitudinal dynamics in these asymmetric collisions.

%%%%%%%%%%%%%%%%%%%%%%  ACKNOWLEDGMENTS}  %%%%% MGS17 version

\begin{acknowledgments}
We thank the staff of the Collider-Accelerator and Physics
Departments at Brookhaven National Laboratory and the staff of
the other PHENIX participating institutions for their vital
contributions.  
We thank Adam Bzdak and Piotr Bo\.{z}ek for providing theoretical
calculations for the suite of collision systems and centralities 
from the wounded-quark model and hydrodynamical model, respectively.
We acknowledge support from the
Office of Nuclear Physics in the
Office of Science of the Department of Energy,
the National Science Foundation,
Abilene Christian University Research Council,
Research Foundation of SUNY, and
Dean of the College of Arts and Sciences, Vanderbilt University
(U.S.A),
Ministry of Education, Culture, Sports, Science, and Technology
and the Japan Society for the Promotion of Science (Japan),
Conselho Nacional de Desenvolvimento Cient\'{\i}fico e
Tecnol{\'o}gico and Funda\c c{\~a}o de Amparo {\`a} Pesquisa do
Estado de S{\~a}o Paulo (Brazil),
Natural Science Foundation of China (People's Republic of China),
Croatian Science Foundation and
Ministry of Science and Education (Croatia),
Ministry of Education, Youth and Sports (Czech Republic),
Centre National de la Recherche Scientifique, Commissariat
{\`a} l'{\'E}nergie Atomique, and Institut National de Physique
Nucl{\'e}aire et de Physique des Particules (France),
Bundesministerium f\"ur Bildung und Forschung, Deutscher Akademischer
Austausch Dienst, and Alexander von Humboldt Stiftung (Germany),
J. Bolyai Research Scholarship, EFOP, the New National Excellence
Program ({\'U}NKP), NKFIH, and OTKA (Hungary),
Department of Atomic Energy and Department of Science and Technology
(India),
Israel Science Foundation (Israel),
Basic Science Research Program through NRF of the Ministry of
Education (Korea),
Physics Department, Lahore University of Management Sciences (Pakistan),
Ministry of Education and Science, Russian Academy of Sciences,
Federal Agency of Atomic Energy (Russia),
VR and Wallenberg Foundation (Sweden),
the U.S. Civilian Research and Development Foundation for the
Independent States of the Former Soviet Union,
the Hungarian American Enterprise Scholarship Fund,
the US-Hungarian Fulbright Foundation,
and the US-Israel Binational Science Foundation.

\end{acknowledgments}

%\clearpage

%\bibliography{ppg221x1}

%merlin.mbs apsrev4-1.bst 2010-07-25 4.21a (PWD, AO, DPC) hacked
%Control: key (0)
%Control: author (0) dotless jnrlst
%Control: editor formatted (1) identically to author
%Control: production of article title (0) allowed
%Control: page (1) range
%Control: year (0) verbatim
%Control: production of eprint (0) enabled
%
 
\end{document}